\documentclass[aps,twocolumn,nofootinbib]{revtex4-1}

\usepackage{graphicx}
\usepackage[body={6.5in, 9in}, right=1in, top=1in]{geometry}
\usepackage{amssymb}
\usepackage{amsmath} 

\newcommand\ee{\end{equation}}
\newcommand\be{\begin{equation}}
\newcommand\eea{\end{eqnarray}}
\newcommand\bea{\begin{eqnarray}}
\newcommand{\sfrac}[2]{{\textstyle\frac{#1}{#2}}}
\newcommand\di{\partial}
\newcommand\mpl{M_{\rm Pl}}

\begin{document}

\title{The squeezed limit of the solid inflation three-point function} 

\author{Solomon Endlich}
\affiliation{Physics Department and Institute for Strings, Cosmology, and Astroparticle Physics,\\
  Columbia University, New York, NY 10027, USA}
\author{Bart Horn}
\affiliation{Physics Department and Institute for Strings, Cosmology, and Astroparticle Physics,\\
  Columbia University, New York, NY 10027, USA}
\author{Alberto Nicolis}
\affiliation{Physics Department and Institute for Strings, Cosmology, and Astroparticle Physics,\\
  Columbia University, New York, NY 10027, USA}
\author{Junpu Wang}
\affiliation{Physics Department and Institute for Strings, Cosmology, and Astroparticle Physics,\\
  Columbia University, New York, NY 10027, USA}

\begin{abstract}
The recently proposed model of `solid inflation' features a peculiar three-point function for scalar perturbations with an anisotropic, purely quadrupolar, squeezed limit. We confirm this result as well as the overall amplitude of the three point-function via an extremely simple computation, where we focus on the squeezed limit from the start and follow the standard logic adopted in deriving the consistency relations. Our system {\em violates} the consistency relations, but in the squeezed limit the three-point function can still be traded for a background-dependent two-point function,  which is immediate to compute. Additionally, we use these simple methods to derive some {new} results -- namely, certain squeezed limits of the three-point correlators involving vector and tensor perturbations as well.

\end{abstract}

\maketitle

\section{Introduction}

Solid inflation is a cosmological model where primordial inflation is driven by a peculiar solid \cite{ENW}. 

In field theory terms, a generic solid can be described via three scalar fields $\phi^I(x)$ with background (or equilibrium) values that depend on the spatial coordinates but not on the time \cite{DGNR};
\be  \label{eq}
\langle \phi^I(x)\rangle = x^I \; .
\ee
Their time-independence survives even when the solid is placed in an expanding FRW universe and gets physically stretched by the Hubble expansion, provided  $\vec x$ is now identified with the FRW comoving coordinates. Of course the solids to which we are accustomed break when stretched by even a small fraction of their original size; however, as long as the solid does not break, the configuration above is a consistent solution that corresponds to a solid physically expanding at the same rate as the universe.
What makes solid inflation possible in the first place is the existence of a field theory for the solid such that neither breaks down---the solid nor the theory---when the system is stretched by many $e$-folds.

This feature can be guaranteed by an approximate {\em internal} dilation symmetry \cite{ENW}
\be \label{dilation}
\phi^I \to \lambda \phi^I  \; ,
\ee
which also guarantees that the `slow-roll' condition $\rho+ p \ll \rho$ is met, thus implying that such a solid can indeed drive a near exponential phase of inflation.
This symmetry is of course not realized in everyday solids---their dynamics are {\em very} sensitive to dilations---and as a consequence they have $\rho+ p \simeq \rho$, which makes them useless for inflationary purposes.

Such an approximate dilation symmetry should supplement the exact internal symmetries obeyed by all homogeneous and isotropic solids, i.e.~constant shifts and rotations \cite{DGNR}
\be
\phi^I \to \phi^I + a^I \; , \qquad \phi^I \to SO(3) \cdot \phi^I   \; .
\ee
The existence of these symmetries is crucial for recovering physical homogeneity and isotropy of the background solution as well as of the dynamics of perturbations, which are formally broken by the equilibrium configuration \eqref{eq}.

Notice that this unusual  breaking pattern of spacetime symmetries never involves breaking {\em time} translations. It is thus natural to expect that the dynamics of cosmological perturbations in this model do not fit into the standard parameterization provided by the effective field theory of inflation developed in \cite{CCFKS}, which identifies adiabatic perturbations with the Goldstone bosons of spontaneously broken time translations \cite{CLNS}. In fact, solid inflation yields physical predictions for cosmological observables that have no counterparts in the effective field theory of inflation, most notably, the absence of adiabatic perturbations during inflation, and a peculiar three-point function for scalar perturbations.  The latter feature will be the focus of our paper.

Perhaps one of the most relevant features of the solid inflation three-point function is its drastic violation of the standard single-field consistency relation \cite{M, CZ} for  the curvature perturbation $\zeta$, 
\be \label{consistency}
\langle \zeta _{\vec q \to 0}\zeta_{\vec k} \zeta_{-\vec k} \rangle' \simeq -(n_S-1) P_{\zeta}(q) P_{\zeta}(k) \; ,
\ee
 which demands that the three-point function factorize and be suppressed by the scalar tilt $(n_S -1) \ll 1$ in the so-called squeezed limit.
In solid inflation, the three-point function for the curvature perturbation in the same limit is claimed to reduce to \cite{ENW}
\be \label{solid squeezed}
\langle \zeta _{\vec q \to 0}\zeta_{\vec k} \zeta_{-\vec k} \rangle' \simeq -\frac{5 }{36}\,\frac{F_Y}{F}\, \frac{ H^4\left(1-3 \cos^2 \theta \right) }{\epsilon^3 c_L^{12}M_{\rm Pl}^4} \,  \frac{1}{q^3 k^3} \; ,
\ee
where $\theta$ is the angle between $\vec q$ and $\vec k$, $F_Y$ and $F$ are free parameters of the solid Lagrangian, and $c_L$ is the speed of longitudinal phonons. We have kept only the leading order in the slow-roll expansion, and in both expressions the prime denotes that a momentum-conserving $(2\pi)^3 \delta^3$ has been removed. 
By using the explicit form for the power spectrum (also to leading order in slow roll\footnote{Note that for this expression to be valid, we must assume not only that the slow-roll parameters are small, but that the slow-roll parameters times the number of $e$-foldings are small.  For a more involved example where the first subleading piece needs to be kept, see the Appendix.}) \cite{ENW}
\be \label{spectrum}
P_{\zeta}(k) = \frac{H^2}{4 \epsilon c_L^5 M_{\rm Pl}^2} \, \frac{1}{k^3} \; ,
\ee
we see that the three-point function above can be written as
\begin{align}
\langle \zeta _{\vec q \to 0}\zeta_{\vec k} \zeta_{-\vec k} \rangle' \simeq & -\frac{20 }{9}\frac{F_Y}{F} \frac{1}{\epsilon c_L^2}  \times\left(1-3 \cos^2 \theta \right) \nonumber \\
& \times P_{\zeta}(q)P_{\zeta}(k)  \; . \label{solid squeezed 2}
\end{align}
This violates the consistency relation \eqref{consistency} in two directions. First, the angular dependence is that of a pure quadrupole, whereas eq.~\eqref{consistency} has no angular dependence whatsoever, that is, it is a pure monopole\footnote{The quadrupolar squeezed limit also appears in the $f(\phi)F^2$ model introduced in \cite{Barnaby:2012tk}.}. Second, once the two spectra are factored out, the overall amplitude of the three-point function is not constrained to be small, nor proportional to the scalar tilt. In fact, the two parameters $F$ and $F_Y$ can in principle be of the same order of magnitude, and the scalar tilt in solid inflation has a contribution $(n_S-1) \supset 2 \epsilon c_L^2$, so the overall size of the three-point function in the squeezed limit can be as large as {\em one over} the scalar tilt \cite{ENW}.
 
Notice that solid inflation is a  model with a {\em single scalar} mode: under the unbroken rotation group, the original scalars' perturbations decompose into one scalar and the two polarizations of a transverse vector. It is thus at first glance surprising that solid inflation can violate the consistency relations, which are supposed to hold for {\em all} ``single-field" models.
At the formal level, however, there is no contradiction, since the peculiar symmetry breaking pattern of solid inflation implies that there is no gauge in which the matter fluctuations are set to zero and the only scalar mode is parameterized by $\zeta $ in the usual manner $g_{ij} \propto e^{2 \zeta} $.
And yet, given that the full computation yielding  \eqref{solid squeezed} is quite involved, it would be useful to have an independent and simpler confirmation of this unusual squeezed limit behavior.
It is the purpose of this note  to provide such a check.

\section{Background field methods} \label{BFM}
The usual derivations \cite{CZ, CFKS} of the consistency relations rely on two main steps. The first involves re-expressing a squeezed limit three-point function first as a nested correlation function, and then as a product of two two-point functions. Schematically, if $\zeta_1$ is a very long-wavelength mode, and $\zeta_2$ and $\zeta_3$ are much shorter, one has
\be \label{nested}
\langle \zeta_1(x) \zeta_2(y) \zeta_3(z) \rangle = \big \langle \zeta_1(x) \, \langle \zeta_2(y) \zeta_3(z) \rangle_{\zeta_1} \big \rangle
\ee
where $\langle \dots \rangle_{\zeta_1}$ denotes a correlation function in the presence of a background field.
Such a formula can be motivated in the following way. For modes that are well outside the horizon at some given time $t$, one can neglect quantum effects and think of the modes as classical stochastic variables. At fixed time $t$, there will be some classical probability distribution functional $P[\zeta]$ for the field's {\em spatial} configuration $\zeta(\vec x)$, which one can formally use in a {\em purely spatial} functional integral to express equal-time correlation functions.
In Fourier space:
\be \label{spatial path}
\langle \zeta_{\vec k_1} \dots \zeta_{\vec k_N} \rangle = \int  [D\zeta] \, P[\zeta]  \, \zeta_{\vec k_1} \dots \zeta_{\vec k_N} \; .
\ee
If one of the modes---say $\zeta_{\vec k_1}$---is much longer than the others, one can perform the path integral in two steps, by first integrating  over the short modes with $k \gg k_1$ in the presence of given long modes, and then integrating over these background long modes at the end:
\begin{align}
\int  [D\zeta] \,  &  P[\zeta, t] \,   \zeta_{\vec k_1} \dots \zeta_{\vec k_N} \to   \\
& \int  [D\zeta_\ell] \zeta_{\vec k_1}   \int [D\zeta_s] \,   P[\zeta_\ell + \zeta_s] \zeta_{\vec k_2} \dots \zeta_{\vec k_N} \; . \nonumber
\end{align}
The integral over the short modes yields the background-dependent correlation function
\be
\langle \zeta_{\vec k_2} \dots \zeta_{\vec k_N} \rangle_{\zeta_\ell}
\ee
times an overall normalization factor, which is nothing but the reduced probability functional for the long modes only:
\be
P_{\rm eff}[\zeta_\ell] \equiv \int [D\zeta_s] \,   P[\zeta_\ell + \zeta_s] \; .
\ee
Then the original equal-time correlation function can be rewritten as
\be \label{nested formal}
\langle \zeta_{\vec k_1} \dots \zeta_{\vec k_N} \rangle = \int  [D\zeta_\ell] P_{\rm eff}[\zeta_\ell] \,  \zeta_{\vec k_1}  \langle \zeta_{\vec k_2} \dots \zeta_{\vec k_N} \rangle_{\zeta_\ell} \; , 
\ee
which is what we actually mean by expressions like \eqref{nested}.

Then---the argument goes---one can Taylor-expand the short-mode correlation function in powers of the background field,
\begin{align}
\langle \zeta_{\vec k_2} \dots \zeta_{\vec k_N} \rangle_{\zeta_\ell} & = \langle \zeta_{\vec k_2} \dots \zeta_{\vec k_N} \rangle \big|_{\zeta_\ell=0} \\
&  + \zeta_\ell *\frac{\delta}{\delta \zeta_\ell}\langle \zeta_{\vec k_2} \dots \zeta_{\vec k_N} \rangle \big|_{\zeta_\ell=0} + \dots \nonumber \; ,
\end{align}
where `$*$' denotes a purely spatial convolution.
The zeroth order term does not contribute to \eqref{nested formal}, because $\zeta$ has vanishing expectation value. The leading contribution thus comes from the first order term, which yields
\be \label{factorized}
\langle \zeta_{\vec k_1} \dots \zeta_{\vec k_N} \rangle \simeq \langle \zeta_{\vec k_1} \zeta_\ell \rangle * 
\frac{\delta}{\delta \zeta_\ell}\langle \zeta_{\vec k_2} \dots \zeta_{\vec k_N} \rangle \big|_{\zeta_\ell=0} \; .
\ee

So  far the derivation is very general. For one thing, we have assumed that all the modes can be treated as classical stochastic variables, but in fact these results make perfect sense beyond the semiclassical limit  \cite{GHN, ABG, HHK}. Since all the $\zeta$ operators inside our equal-time correlation functions commute with one another, one can just interpret the probability distribution functional $P$ as the square of the wave functional at time $t$, and then the purely spatial path-integrals above are nothing but the usual expressions for quantum mechanical equal-time expectation values in the Schr\"odinger representation, i.e.~$\langle {\cal O}(q) \rangle = \int dq \, {\cal O} (q) \, |\psi(q)|^2$.
\footnote{The wave-functional at time $t$ admits a path-integral representation from $t' = -\infty $ to $t' = t$. The {\em square} of the wave-functional involves two such path-integrals. Our formula \eqref{spatial path} for computing correlators thus reduces to the standard in-in path-integral representation of equal-time correlators.}
Moreover, since no special property of $\zeta$ has been used so far, all the manipulations above can be generalized to non-minimal models---for instance to accommodate more fields. In a generic model, if one is able to compute the first-order dependence on background fields for an $(n-1)$-point correlation function, one immediately has the squeezed limit of the correspoding $n$-point correlation function.

What is special about the standard single-field inflation case? In this scenario a very long wavelength $\zeta$ is a pure gauge mode to zeroth order in gradients, and can thus be set to zero via a rescaling of the spatial coordinates. One can then express the derivative with respect to~$\zeta_\ell$ in \eqref{factorized} as a derivative with respect to~{\em scale}, which in the case of the two-point function is proportional to the tilt, thus  ending up with consistency relations like \eqref{consistency}. This is the second main step we alluded to above, and is not available in solid inflation, because a long wavelength curvature perturbation $\zeta$ is {\em not} equivalent to a rescaling of spatial coordinates. 
Even so, nothing prevents us from following the standard consistency-relation logic all the way to eq.~\eqref{factorized}, which for the three-point function case  reads simply
\be
 \label{factorized2}
\langle \zeta_{\vec k_1} \zeta_{\vec k_2} \zeta_{\vec k_3} \rangle \simeq \langle \zeta_{\vec k_1} \zeta_\ell \rangle * 
\frac{\delta}{\delta \zeta_\ell}\langle \zeta_{\vec k_2}  \zeta_{\vec k_3} \rangle \big|_{\zeta_\ell=0} \; .
\ee


\section{Scalar three-point function in solid inflation}\label{scalars}
Our first task then is to compute the two point function of two high momentum modes in the presence of some long wavelength  perturbation. In practice, in solid inflation it is more convenient to first compute the correlation functions of 
the phonon field $\pi^I \equiv \phi^I -x^I$ in spatially flat gauge, and then convert them to correlation functions of $\zeta$ via the linear relation \cite{ENW}
\be \label{zeta}
\zeta = \sfrac13 \vec \nabla \cdot \vec \pi \; .
\ee
Since we are interested in taking the derivative of the short-mode two-point function with respect to $\zeta_\ell$ and then evaluating it at $\zeta_\ell=0$, there are going to be two pieces from the full action that will give us relevant contributions:\ the longitudinal phonon quadratic action \cite{ENW} 
\begin{align}
S^{(2)}  = & \, \mpl^2 \int \! dt \frac{d^3 k}{(2 \pi)^3}  \, a^3 \bigg\{ \frac{ k^2/3}{1+k^2/3a^2\epsilon H^2}\big|\dot{\pi} +\epsilon H\pi\big|^2 \nonumber
\\ & -\epsilon H^2c_L^2 \, k^2 \big| \pi  \big|^2 \bigg\} \;, \label{quadratic}
\end{align}
where $\pi \equiv \hat k \cdot \vec \pi$ parameterizes the longitudinal mode, 
and the cubic action \cite{ENW}
\begin{align}
S^{(3)} = &  \, \mpl^2\int \! d^4 x \, a^3   H^2\frac{F_Y}{F}  \Big\{\sfrac{7}{81}(\di_i \pi^i)^3-\sfrac{1}{9}\di_i  \pi^i \di_j\pi^k\di_k\pi^j  \nonumber \\
& -\sfrac{4}{9}\di_i \pi^i \di_j\pi^k\di_j\pi^k +\sfrac{2}{3}\di_j\pi^i \di_j\pi^k\di_k\pi^i\Big\}  \; . \label{cubic}
\end{align}
The complicated non-local structure in \eqref{quadratic} comes from having integrated out $N$ and $N^i$ via the constraint equations. In principle this yields extra contributions (some of which are also non-local) to the cubic action as well, but one can check that the terms we have kept are the leading ones in the slow-roll expansion \cite{ENW}.

In the presence of a very long wavelength background field,
 the cubic action expanded to first order in the background gives rise to a correction to the quadratic action for the short modes. Restricting to the longitudinal phonons only and using $ \di^i \pi_\ell^j=3 \,  \hat{k}_\ell^i \hat{k}_\ell ^j \, \zeta_\ell $,
this is  simply
\begin{align}
\Delta S^{(2)}  = & - M_{\rm Pl}^2 \int \! d^4 x \, a^3 H^2 \frac{8}{9}\frac{F_Y}{ F} \, \zeta_\ell \nonumber  \\
& \times 
 \left(\delta^{ij}-3 \hat{k}_\ell^i \hat{k}_\ell^j \right) \di_i \pi^k \di_j \pi^k \;, \label{Delta S2}
\end{align}
where we have freely integrated by parts and used $\di_i \pi_j = \di_j \pi_i$, which is appropriate for longitudinal modes. We can already see the emergence of the  quadrupole structure discussed above.

To lowest order in slow roll, all the parameters ($H, \epsilon, c_L, F, F_Y$) appearing in $S^{(2)}$ and $\Delta S^{(2)}$ can be treated as constant. Moreover, at the freeze-out time for the short modes (which, as usual, roughly corresponds to the time when the correlation functions get their dominant contributions), the long background mode is well outside the horizon and is thus approximately constant in time, with a relative time dependence of order $\epsilon$ \cite{ENW}. We thus see that, to lowest order in slow-roll, the only effect of a long wavelength background is  a direction-dependent correction to the longitudinal phonon speed:
\be
c_L^2 \to c_L^2+ \frac89 \frac{F_Y}{F}\frac{1}{ \epsilon} (1-3 \cos ^2\theta) \zeta_\ell \; \equiv \; \tilde{c}_L^2 \; ,
\ee
where $\theta$ is the angle between the momenta of the short and long modes.
The short-mode two-point function associated with $S^{(2)}+\Delta S^{(2)}$ is therefore nothing but the original one,\ eq.~\eqref{spectrum}, with  $c_L \to \tilde{c}_L$. We thus get that to zeroth order in $\vec k_\ell$, the relevant functional derivative with respect to the background is simply
\footnote{For functional derivatives in Fourier space, we are using the convention
$\frac{\delta}{\delta f(\vec q_1) } \, f(\vec q_2)=(2\pi)^3 \delta^3(\vec q_2 - \vec q_1)$,
 which goes well with the standard $d^3k/(2\pi)^3$ integration measure.}
\begin{align}
\frac{\delta}{\delta \zeta_\ell}\langle \zeta_{\vec k_2}  \zeta_{\vec k_3} \rangle \big|_{\zeta_\ell=0} & = (2\pi)^3 \delta^3(\vec k_2 + \vec k_3 - \vec k_\ell)  \nonumber \\
& \times
\frac{\di}{\di c_L^2}\langle \zeta_{\vec k_2}  \zeta_{\vec k_3} \rangle' \cdot \frac{\di \tilde{c}_L^2}{\di \zeta_\ell}
\end{align}
with
\be
\frac{\di}{\di c_L^2}\langle \zeta_{\vec k_2}  \zeta_{\vec k_3} \rangle' \cdot \frac{\di \tilde{c}_L^2}{\di \zeta_\ell} = - \frac{20}{9}\frac{F_Y}{F} \frac{1}{\epsilon c_L^2} (1-3 \cos ^2\theta) \langle \zeta_{\vec k_2}  \zeta_{\vec k_3} \rangle ' \nonumber
\ee
Plugging this into \eqref{factorized2}, performing the trivial convolution over the delta functions, and factoring out a total-momentum delta function, one gets precisely the desired result \eqref{solid squeezed 2}.
This provides an independent check of the three-point function computation of \cite{ENW}, and confirms the peculiar size and angular dependence that emerge in the squeezed limit \footnote{Strictly speaking, the calculation in \cite{ENW} was performed in the ``not-so-squeezed'' regime where $q^2/k^2 > \epsilon$ to simplify the analytic expression.  A more thorough treatment allows us to push the original calculation to $q \to 0$, however, and we have checked that the results match the squeezed limits reported here.}. 

\section{Symmetry constraints on the angular dependence}
It is interesting to ponder what the fundamental reason behind the purely quadrupolar angular dependence may be 
\footnote{We thank Enrico Pajer for posing the question in the first place.}.
At first sight, a monopole contribution seems to be allowed by all the symmetries and by the derivative structure of the cubic interactions \eqref{cubic}. Yet when all contributions are added up in \eqref{Delta S2}, one is left with no monopole.  The symmetry reason behind these cancellations is that a monopole would violate the approximate dilation symmetry \eqref{dilation}. To see this, consider the generic structure for  $\Delta S^{(2)}$ one gets by using one power of a general background field $\pi^0$ in $S^{(3)}$:
\be \label{generic Delta S2}
S^{(3)} \to \Delta S^{(2)} \propto \int T^{ij,kl,mn} \di_i \pi^0_j \di_k \pi^s_l \di_m \pi^s_n \; .
\ee 
The $T$ tensor can be taken to be symmetric  under the exchange of any two of the $(ij)$, $(kl)$, and $(mn)$   pairs (if all the fields are longitudinal, it can also be taken to be symmetric within each pair).
The action of \eqref{dilation} on the phonon field is
\be
\vec \pi \to \vec \pi + \omega \, \vec x + \dots  \; , \qquad \omega \equiv \lambda-1
\ee
where, for simplicity, we are restricting to infinitesimal $\omega$, and we are keeping the leading order only.
At the level of derivatives of $\vec \pi$, the infinitesimal symmetry is
$ \di_i \pi_j \to \di_i \pi_j + \omega \, \delta_{ij}$, which 
 implies that a background field with $\di_i \pi^0_j \propto \delta_{ij}$ is physically trivial, i.e.\ equivalent to no background at all, since it can be `undone' via a symmetry transformation. 
Then, using such a background in \eqref{generic Delta S2} should yield zero, which means that the $(ij)$ trace of $T \cdot (\di \pi) (\di \pi)$ vanishes. In particular, for longitudinal short modes with momentum $\vec k_s$, traceleness in $(ij)$ implies
\be \label{traceless T}
T^{ij, kl, mn} \di_k \pi^s_l \di_m \pi^s_n \propto (\delta^{ij} - 3 \, \hat k_s^i \hat k_s^j ) \; . 
\ee
Contracting this now with a physical background of momentum $\vec k_\ell$ yields our quadrupole structure.
These considerations make it clear that beyond lowest order in slow-roll we do expect to get a monopole signal in the squeezed limit of the three-point function, since  slow-roll corrections break the dilation symmetry \eqref{dilation}.

We can generalize this argument to accommodate vector perturbations as well. This will provide us with a useful check for the results of the next section. Without specializing to scalar or vector perturbations, but keeping instead generic polarizations, from \eqref{generic Delta S2} we have that the tensor structure of a generic squeezed three-point function
is
\begin{align}
\langle \pi^\ell _{\lambda_\ell, \vec k _\ell} \pi^s _{\lambda_1, \vec k_s} \pi^s_{\lambda_2, -\vec k_s }\rangle & \propto 
T^{ij, kl, mn} \times \\
& \hat k_\ell^i \hat \varepsilon^j _ {\lambda_\ell} \, \hat k_s^k \hat \varepsilon^l _ {\lambda_1} \, \hat k_s^m \hat \varepsilon^n _ {\lambda_2}
 \nonumber \; ,
\end{align}
where the $\lambda$'s denote the polarizations, and the $\hat \varepsilon$'s the corresponding polarization vectors.
The dilation-invariance argument above still tells us that
\be \label{combination}
T^{ij, kl, mn}  \, \hat k_s^k \hat \varepsilon^l _ {\lambda_1} \, \hat k_s^m \hat \varepsilon^n _ {\lambda_2}
\ee
is traceless, but now this combination  depends on the polarization vectors as well, and so we cannot rewrite it as simply as in eq.~\eqref{traceless T}. If the short modes are scalars there is no problem of course, because their polarization vectors are aligned with $\hat k_s$, but for  vector short modes, to get a result that only depends on the momentum we have to {\em sum} (or average) over polarizations
\footnote{We are implicitly using that the transverse polarization vectors provide an orthonormal basis for  the plane orthogonal to $\vec k_s$, that is $\sum_\lambda \hat \varepsilon_\lambda^i \hat \varepsilon_\lambda^j = \delta^{ij} - \hat k_s^i \hat k_s^j$.}. 
So, in general we can write
\be
{\sum_{\lambda_1} } \,   T^{ij, kl, mn}  \, \hat k_s^k \hat \varepsilon^l _ {\lambda_1} \, \hat k_s^m \hat \varepsilon^n _ {\lambda_1} \propto (\delta^{ij} - 3 \, \hat k_s^i \hat k_s^j ) \; ,
\ee
where it is understood that the sum runs over all values that are appropriate for the perturbations in question---i.e.~no sum for scalars, and a sum over the two transverse polarizations for  vectors. We thus get a sum rule for the angular dependence of a (fairly) generic squeezed three-point function
\begin{align}
{\sum_{\lambda_s} }  \langle \pi^\ell _{\lambda_\ell, \vec k _\ell} & \pi^s _{\lambda_s, \vec k_s} \pi^s_{\lambda_s, -\vec k_s }\rangle  \propto   \label{sum rule} \\
& (\hat k_\ell \cdot \hat \varepsilon _ {\lambda_\ell}) - 3  (\hat k_\ell \cdot \hat k_s) (\hat \varepsilon _ {\lambda_\ell} \cdot \hat k_s) \nonumber \; ,
\end{align}
of which our quadrupolar  result for the purely scalar three-point function is just a special case.

We say `fairly generic' because the sum rule above does not apply to squeezed limits in which the two short modes are a scalar {\em and} a vector. For those, we can replace $\hat \varepsilon_{\lambda_1}$ in eq.~\eqref{combination} with $\hat k_s$. Then, the traceleness of such a combination and its involving only one transverse polarization vector imply that it can be rewritten as
\begin{align}
T^{ij, kl, mn}  \, \hat k_s^k \hat k_s^l \hat k_s^m \, \hat \varepsilon^n _ {\lambda_2} & \propto (\delta^{ij} - 3 \, \hat k_s^i \hat k_s^j ) \\
& + \beta \, \hat k_s^i \hat \varepsilon^j _ {\lambda_2} 
+ \gamma \, \hat \varepsilon^i_ {\lambda_2}   \nonumber 
 \hat k_s^j \; ,
\end{align}
where $\gamma$ and $\beta$ are two unconstrained constants. This still implies a constraint on the angular structure of the three-point function in question, but one that is considerably looser than the one above, which involves no free parameters.

\section{Three-point correlators with vector modes}
The simple computational method discussed in sect.~\ref{scalars} is of course very general, and not limited to our model, nor, within our model, to scalars only. For instance, we can calculate the squeezed limit of the scalar-vector-vector three-point function. For simplicity, let us take the scalar mode to be the long wavelength one, and therefore, using the logic outlined above, we have that
\be
\label{scalar-vector-vector-factorized}
\langle \zeta_{ q} \, \pi_{T, k_1}^{i} \pi_{T, k_2}^{j}  \rangle \simeq \langle \zeta_{ q} \zeta_\ell \rangle * 
\frac{\delta}{\delta \zeta_\ell}\langle  \pi_{T, k_1}^{i} \pi_{T, k_2}^{j} \rangle \big|_{\zeta_\ell=0} \; ,
\ee
where the vector $\vec \pi_T$ is the transverse phonon field in spatially flat gauge. In Fourier space its quadratic action is given by 
\begin{align}
S^{(2)}_T  = & \, \mpl^2 \int \! dt \frac{d^3 k}{(2 \pi)^3}  \, a^3 \bigg\{\frac{ k^2/4}{1+k^2/4a^2\epsilon H^2} \, \big| \dot{\pi}_T^i \big|^2 \nonumber
\\ & -\epsilon H^2 c_T^2 \, k^2 \big|\pi_T^i  \big|^2 \bigg\}  \; , \label{Tquadratic}
\end{align}
which yields a spectrum of super-horizon modes
\be \label{vector spectrum}
P_T(k) = \frac{9 H^2}{4 \epsilon c_T^5 M_{\rm Pl}^2} \, \frac{1}{k^5}=9 \, \frac{c_L^5}{c_T^5} \, \frac{P_\zeta(k)}{k^2}
\ee
for each of the two transverse polarizations of $\vec \pi_T$ \cite{ENW}.
The different $k$-dependence from the scalar two-point function just arises from the $k$-dependent relation between $\zeta$ and $\vec \pi$, eq.~\eqref{zeta}.

Inserting one long wavelength scalar mode and two vector modes into the cubic action (\ref{cubic}), and freely integrating by parts while taking advantage of the identities $ \di^i \pi_L^j = \di^j \pi_L^i$ and $\di_i \pi_T^i=0$, we have a correction to the quadratic action of the vector perturbations:
\begin{align}
\Delta S_T^{(2)}  = & - M_{\rm Pl}^2 \int \! d^4 x \, a^3 H^2 \frac{2}{3} \frac{F_Y}{ F} \, \zeta_\ell \, \di_a \pi_T^b \di_c \pi_T^d \nonumber  \\
& \times 
 \left(2\delta^{ac}\delta^{bd}-3\delta^{bd} \hat{k}_\ell^a \hat{k}_\ell^c -3\delta^{ac} \hat{k}_\ell^b \hat{k}_\ell^d \right)  \;. \label{Delta ST2}
\end{align}
Once again, just as above, working to lowest order in slow roll, the only effect of a long wavelength background is a direction-dependent (and polarization dependent) correction to the transverse phonon speeds.
%
In fact, the last term generically induces a direction-dependent mixing between the two transverse polarizations. To remove this mixing, one could proceed by diagonalizing this last term by a judicious choice of the polarization vectors. However, we can instead proceed in a quicker (or lazier) way by simply recognizing that in the full three-point function---of which we are calculating just the squeezed limit---the vector structure comes from the cubic vertex above, as so it follows that 
\begin{align}
\langle \zeta _{\vec q \to 0} \pi_{\lambda_1, \vec k} \, \pi_{\lambda_2, -\vec k} \rangle &\propto \big[ -3 (\hat{q} \cdot \hat{\varepsilon}_{\lambda_1})(\hat{q} \cdot \hat{\varepsilon}_{\lambda_2})\nonumber \\
&+(2-3 \cos^2\theta)  (\hat{\varepsilon}_{\lambda_1}\cdot  \hat{\varepsilon}_{\lambda_2})
\big] \; , \label{structure of long L TT}
\end{align}
where we have expressed the vector field as a sum over the two transverse polarizations:
\be
\vec{\pi}_T(\vec k)=\sum_{\lambda} \hat{\varepsilon}_{\lambda}(\vec k) \, \pi_{\lambda}( \vec k) \; .
\ee
\noindent where the polarization vectors are normalized so that $\hat \varepsilon_\lambda (\vec k) \cdot \hat \varepsilon_{\lambda'}(-\vec{k}) = \delta_{\lambda \lambda'}$.  

Now, we determine the overall coefficient by doing a much more restricted calculation. Choosing the polarization of both the vector modes to be perpendicular to the momentum of the long mode,
\be \label{eps perp}
\hat{\varepsilon}_{\lambda_1}, \hat{\varepsilon}_{\lambda_2} \to \hat \varepsilon_\perp \propto \hat k_\ell \times \vec k \; ,
\ee
eq.~(\ref{Delta ST2}) becomes a simple shift in the phonons' speed $c_T \to \tilde{c}_{T , \perp}$ as before, where
\be
\tilde{c}^2_{T , \perp}  \equiv  c_T^2 +\frac{2}{3}\frac{F_Y}{F}\frac{1}{ \epsilon}\left(2-3 \cos ^2 \theta \right)\zeta_\ell \; .
\ee
Moving forward as we did above, we can express
\begin{align}
\frac{\delta}{\delta \zeta_\ell}\langle  \pi_{\perp, k_1} \pi_{\perp, k_2} \rangle \big|_{\zeta_\ell=0}& = (2\pi)^3 \delta^3(\vec k_1 + \vec k_2 - \vec k_\ell)  \nonumber \\
 \times
 \frac{\di}{\di c_{T }^2}&\langle  \pi_{\perp, k_1} \pi_{\perp, k_2}  \rangle' \cdot \frac{\di c_{T , \perp}^2}{\di \zeta_\ell} \; .
\end{align}
Using the  two-point function of the vector modes \eqref{vector spectrum} and putting everything together as in \eqref{scalar-vector-vector-factorized} we can compare to (\ref{structure of long L TT}) . This gives us the overall coefficient and the promised three-point function:
\begin{align}
\langle \zeta _{\vec q \to 0} & \pi_{\lambda_1, \vec k} \pi_{\lambda_2, -\vec k} \rangle'  \simeq -\frac{5 }{3}\,\frac{F_Y}{F}\, P_{\zeta}(q) P_{T}(k) \, \frac{1}{\epsilon c_T^2} \,  \nonumber  \\
 \times & \left[(2-3 \cos^2\theta)  (\hat{\epsilon}_{\lambda_1}\cdot  \hat{\epsilon}_{\lambda_2})- 3 (\hat{q} \cdot \hat{\epsilon}_{\lambda_1})(\hat{q} \cdot \hat{\epsilon}_{\lambda_2})\right] \; .
\end{align}
Apart from the obvious differences in the angular (and tensor) structure, this result is suppressed with respect to the purely longitudinal case \eqref{solid squeezed} by a $(c_L/c_T)^7$ factor, which, as discussed in \cite{ENW}, can range from arbitrarily small for $c_L \ll 1$, to roughly $2\%$ for $c_L \simeq 1/3$, $c_T \simeq 1$. Notice that if we average over the two polarizations the angular dependence reduces once again to that of a pure quadrupole, as predicted by our sum rule \eqref{sum rule}.

Similarly, we can compute the vector-scalar-scalar three-point function in the limit of a long vector mode. This is, once again, easy as the coupling of the background long wavelength vector perturbation to the scalar modes gives just a direction-dependent renormalization of the longitudinal speed of sound: $c_L^2 \rightarrow \tilde{c}_L^2$. A nearly identical calculation as above yields:
\begin{align}
\langle \pi _{\lambda, \, \vec q \to 0} \zeta_{ \vec k} \zeta_{ -\vec k} \rangle'  \simeq i q \frac{20 }{9}\,\frac{F_Y}{F}\,& P_{T}(q) P_{\zeta}(k) \, \frac{1}{\epsilon c_L^{2}} \, \nonumber  \\
 \times & (\hat{\varepsilon}_\lambda \cdot \hat{k}) \cos \theta\; ,
\end{align}
which obeys the sum rule \eqref{sum rule}.

The vector-vector-vector three-point function in the squeezed limit is also within reach of these techniques. We proceed as we did for the $\langle \zeta _{\vec q \to 0} \pi_{\lambda_1} \pi_{\lambda_2} \rangle$ correlation function. From the structure of the cubic action it is clear that 
\begin{align}
\langle \pi _{\lambda, \vec q \to 0}& \pi_{\lambda_1, \vec k} \pi_{\lambda_2, -\vec k} \rangle \propto \big[2(\hat{q} \cdot \hat{k})(\hat{\varepsilon}_{\lambda}\cdot \hat{k} )(\hat{\varepsilon}_{\lambda_1}\cdot \hat{\varepsilon}_{\lambda_2} )\nonumber \\
&+(\hat{\varepsilon}_{\lambda_1}\cdot \hat{q})( \hat{\varepsilon}_{\lambda_2} \cdot \hat{\varepsilon}_{\lambda} )+
(\hat{\varepsilon}_{\lambda_2}\cdot \hat{q})( \hat{\varepsilon}_{\lambda_1} \cdot \hat{\varepsilon}_{\lambda} )
\big] \; . \label{structure of squeezed TTT}
\end{align}
Choosing a particularly convenient choice of polarizations one can now fix the overall coefficient as before. 
One can for instance choose the long mode to be polarized in the plane defined by $\hat k_\ell$ and $\hat k$, so that $\hat{\varepsilon}_{\lambda}\cdot \hat{k}=\sin \theta$. Then, one can choose the short modes to be polarized in the orthogonal direction, as in eq.~\eqref{eps perp}, so that $\hat{\varepsilon}_{\perp} \perp \hat{k}_{\ell},\hat{\epsilon}_{\lambda}$. 
For this particular configuration, the propagation speed of the short modes  shifts as
\be
c_T^2 \to \tilde{c}^2_{T,\perp}  \equiv  c_T^2 -\frac{2}{3}\frac{F_Y}{F}\frac{1}{ \epsilon}(i k_\ell)(\cos \theta \sin \theta )\pi_{\lambda, \vec k_\ell} \; .
\ee
We can then calculate the three-point function $\langle \pi _{\lambda, \vec q \to 0} \pi_{\perp, \vec k} \pi_{\perp, -\vec k} \rangle$  and  match with (\ref{structure of squeezed TTT}) in order to determine the overall coefficient. When the dust clears we are left with the desired result:
\begin{align}
\langle \pi _{\lambda, \vec q \to 0}& \pi_{\lambda_1, \vec k} \pi_{\lambda_2, -\vec k} \rangle' \simeq iq \, \frac{5}{6}\,P_T(q) P_T(k) \frac{F_Y}{F} \, \frac{1}{\epsilon c_T^2 }  \nonumber \\ & \times \big[2 \cos  \theta \, (\hat{\varepsilon}_{\lambda}\cdot \hat{k} )(\hat{\varepsilon}_{\lambda_1}\cdot \hat{\varepsilon}_{\lambda_2} )\nonumber \\
&+(\hat{\varepsilon}_{\lambda_1}\cdot \hat{q})( \hat{\varepsilon}_{\lambda_2} \cdot \hat{\varepsilon}_{\lambda} )+
(\hat{\varepsilon}_{\lambda_2}\cdot \hat{q})( \hat{\varepsilon}_{\lambda_1} \cdot \hat{\varepsilon}_{\lambda} )
\big]  \; . \label{squeezed TTT}
\end{align}
Recalling that $\hat q$ and $\hat \varepsilon_\lambda$ are orthogonal, it is a matter of straightforward algebra to check that the sum rule \eqref{sum rule} is obeyed in this case as well.

The careful reader might have noticed that we have computed all the squeezed limits of three-point functions
involving scalars and vectors, apart from those in which the short modes are one vector and one scalar, i.e.
\be
\langle \zeta_{ \vec q} \, \zeta_{ \vec k} \,  \pi^i_{T, -\vec k} \rangle \quad  \mbox{and} \quad \langle \pi^j_{T, \vec q} \, \zeta_{ \vec k} \,   \pi^i_{T, -\vec k} \rangle 
\ee
with $q \to 0$.
The problem with these is that  the background long mode induces a mixing between the scalar and  vector short modes, which makes their mixed two-point function {\em (a)} nonzero, and {\em (b)} not immediate to compute, because it cannot be derived through a simple background-dependent renormalization of a parameter already appearing in the quadratic Lagrangian for these modes. One really has to diagonalize the scalar-vector system in the presence of this mixing and then compute the relevant two-point functions anew, which we leave for future work. Notice that treating the mixing term as a small correction and trying to assess its effect on the scalar-vector two-point function in perturbation theory is not particularly effective from our standpoint: the leading order contribution to our mixed two-point function comes from a diagram with two external legs and one insertion of the mixing vertex, {\em integrated} over the time of this vertex (in the standard $t, \vec k$ representation),
which is not much easier---nor much different, in fact---than the full three-point function computation.

For completeness we collect in the Appendix a number of squeezed three-point functions involving tensor modes as well.

\section{Concluding Remarks}
The techniques we have outlined in this paper are clearly very general, and can be applied to all situations in which {\em (i)} the two-point functions for the perturbations are known, and {\em (ii)} the only effect of a very long-wavelength background field at the relevant time (e.g.\ at the freeze-out time) is a renormalization of the coefficients already appearing in the quadratic Lagrangian for the perturbations. 
In principle these techniques can be extended to higher orders as well: we plan to use them to compute the solid inflation four-point function in the ``triangular'' limit.

We close with a couple of comments on the validity of our methods. First, we have used throughout the paper that to lowest order in the slow-roll expansion the long-wavelength background fields are constant in time. However, this property is not crucial for our methods to be applicable, and these can thus be extended to higher orders in slow-roll. For instance, in the purely scalar case, if we take into account the weak time-dependence of $\zeta_\ell$ in \eqref{Delta S2}, we get a weakly time-dependent renormalization of $c_L$. But since at first order in slow roll $c_L$ is weakly time-dependent to begin with---with rate $s = \dot c_L/(c_L H)$---this can be seen as a $\zeta_\ell$-dependent simultaneous renormalization of  $c_L$ {\em and} $s$, with obvious implications for the first-order-in-slow-roll spectrum of perturbations. 

Second, and perhaps more relevant: even though  the general discussion in sect.~\ref{BFM} makes sense at the quantum mechanical level as well, when we implement these techniques in the subsequent sections  we make implicit use of the assumption that the long background mode is in fact classical. The reason is somewhat subtle, and it deserves to be spelled out in detail.
Recall that all correlators in sect.~\ref{BFM} are equal-time correlators, which admit a purely spatial, equal-time functional integral representation. As we mentioned there, at this level there is no difference between quantum field theory and classical statistical field theory, provided we identify $P = | \Psi|^2 $. The difference between the two then must come from time-evolution. Indeed,  for any given realization at time $t$,  in classical statistical field theory we can evolve it  in time just by using the classical equations of motion. By contrast, in quantum field theory we should integrate over all trajectories with that boundary condition at $t$, with the usual path-integral measure.
This difference is crucial when we are asked to compute a background-dependent short-mode correlator like $\langle \zeta_s \zeta_s \rangle_{\zeta_\ell}$. According to our discussion in sect.~\ref{BFM}, all these modes are evaluated at the same time $t$. In particular, the background $\zeta_\ell$ is only specified at time $t$. Then, in principle, to compute this correlator we have to extrapolate this given background configuration back in time and assess its effect on the build-up of the short-mode correlator, from $-\infty$ to $t$. If the long mode can be treated as classical, this extrapolation just involves the equations of motion. In our case, they taught us that the long mode is constant in time to lowest order in slow-roll. On the other hand, if one wants to go beyond the classical limit and treat the long mode at the quantum level, then this extrapolation back in time involves a path-integral with given future boundary
conditions at $t$, which, although well-defined, clearly complicates the computation substantially.

\vspace{.3cm}

\noindent
{\em Note added---}
Recently, motivated by the violation of the `cosmic no-hair theorem' \cite{Wald} in this model, an interesting paper  \cite{BMPR} computed up to an order-one factor the scalar two point function in the presence of an anisotropic gravitational background.
In perturbation theory, their anisotropic metric perturbation can be interpreted as a zero-momentum tensor mode
\be \label{gamma}
\gamma_{ij}= \text{diag}(2\sigma, -\sigma, -\sigma)\; .
\ee
With this identification, the computation of the scalar two-point function on the anisotropic background is amenable to the techniques spelled out throughout this paper. 
We just need the trilinear action for one tensor  and two scalar modes, which to lowest order in slow-roll reads\footnote{
This can be found by expanding the full action for solid inflation contained in \cite{ENW}. We note that working to lowest order in the slow-roll parameters allows one to neglect $\delta N$ and $N^i$ as they will be of order $\epsilon \pi$.
}:
\begin{align}
S_{\gamma \zeta \zeta}  = & \, M_{\rm Pl}^2 \int d^4 x \, a^3 H^2 \, \frac{F_Y}{F}\big\{\sfrac{8}{9} \gamma_{ij} \, \di^i \pi^j \, \di_k \pi^k \nonumber\\
& -\sfrac{4}{3} \gamma_{ij} \, \di^i \pi^k \di^j \pi^k) \big\} \; ,
\end{align}
where we have used that for scalar modes $\di_i \pi_j $ is a symmetric matrix.
For a very long wavelength background $\gamma$, we can then see immediately that this interaction term is, once again, just a renormalization of the speed of the longitudinal modes $c_L^2 \rightarrow \tilde{c}_L^2$ where
\be
\tilde{c}_L^2=c_L^2+\frac{4}{9} \frac{F_Y}{F} \frac{1}{\epsilon} (\hat{k}^i \hat{k}^j \gamma_{ij}) \; .
\ee
%
%
For the $\gamma_{ij}$ of eq.~\eqref{gamma}, and letting $\theta$ now denote the angle between $\vec{k}$ and the  $\hat x$ direction, this corrects the scalar spectrum as 
\be
P_\zeta(k) \rightarrow P_\zeta(k) \left(1-\sigma \frac{10}{9} \frac{F_Y}{F} \frac{1}{ c_L^2 \epsilon} (3 \cos^2 \theta - 1) \right) \;,
\ee
thus allowing us to compute the order-one factor that was left generic in \cite{BMPR}.

\vspace{.3cm}

\noindent
{\em Acknowledgements---} 
We wish to thank Paolo Creminelli,  Walter Goldberger, Lam Hui, and Matias Zaldarriaga for helpful discussions.  This work is supported in part by NASA under the contract NNX10AH14G and by the Department of Energy under contracts DE-FG02-11ER41743 and DE-FG02-92ER40699.  SE was supported by a National Science Foundation Graduate Research Fellowship.

\appendix*

\section{Tensors}

The cubic part of the solid Lagrangian involving a single tensor can be found by expanding the action in \cite{ENW}:
\begin{align}
\mathcal{L}_{\gamma \pi \pi} = &  \, \mpl^2\, a^3   H^2\frac{F_Y}{F}  \Big\{\sfrac{8}{9}\di_k \pi^k  \gamma_{ij} \di^i \pi^j -\sfrac{2}{3}\gamma_{ij} \di^j\pi^k\di_k\pi^i  \nonumber \\
& -\sfrac{1}{3}\gamma_{ij} \di_j\pi^k\di_j\pi^k -\sfrac{1}{3}\gamma_{ij} \di_k\pi^j\di_k\pi^i\Big\} \; .  \label{cubicappendix}
\end{align}
\noindent This yields the soft limits
\be
\langle \gamma^\lambda_{ \vec q \rightarrow 0} \zeta_{\vec k} \zeta_{-\vec k} \rangle'=  -\frac{10}{9}\, \frac{F_Y}{F}\,P_{\gamma}(q) P_\zeta (k)\,\frac{1}{c_L^2 \epsilon}\,   \big(\hat k ^i \hat k ^j \epsilon^\lambda_{ ij} \big) \; ,
\ee
\begin{align}
\langle \gamma^\lambda_{ \vec q \rightarrow 0} \pi_{\lambda_1, \vec k} \pi_{\lambda_2, -\vec k} \rangle' = & -\frac{5}{6} \,\frac{F_Y}{F}\,P_{\gamma}(q) P_T (k)\,\frac{1}{c_T^2 \epsilon}  \nonumber\\ & \times \epsilon^\lambda_{ ij}\big(\hat k ^i \hat k ^j \varepsilon_{\lambda_1} \cdot \varepsilon_{\lambda_2} + \varepsilon^i_{\lambda_1} \varepsilon^j_{\lambda_2} \big) \; ,
\end{align}

\noindent for the scalar and vector cases, where the polarization tensors are normalized according to $\epsilon^s_{ij}(\vec k) \epsilon^{s'}_{ji}(-\vec k) = 2\delta^{s s'}$, and where the power spectrum is given in \cite{ENW}:
\be
P_{\gamma}(q) = \frac{H^2}{M^2_{\rm Pl}}\frac{1}{q^3} \; .
\ee

We omit the mixed tensor-scalar-vector case, for the same reasons as before, as well as the relations where the two-point function for the short modes contains a single tensor.  Since $P_{\gamma}(k)$ does not depend on speed $c^2_T$ to leading order in slow-roll, in order to find the correlators $\langle \zeta \gamma \gamma \rangle$, $\langle \pi_T \gamma \gamma \rangle$, $\langle \gamma \gamma \gamma \rangle$ we need to keep the next to leading corrections in the expression in \cite{ENW}:
\begin{align}
P_{\gamma}(k) = & \, \frac{H_c^2}{M^2_{\rm Pl}}\frac{(k/a H)^{8 c^2_{T} \epsilon_c /3}}{(k/a_c H_c)^{\epsilon_c}}\frac{1}{k^3} \nonumber \\
&\sim \frac{H_c^2}{M^2_{Pl}}\frac{1}{k^3}\left\{1 + \frac{8 c^2_{T} \epsilon_c}{3}\log\left(\frac{k}{a H}\right) - \epsilon_c \log\left(\frac{k}{a_c H_c}\right)\right\}
\label{correctedpower}
\end{align}
\noindent where the subscript denotes the value of the parameter at some fiducial time (e.g.\ at the horizon crossing time for the longest observable mode), and the approximation is appropriate when the logarithm is large (typically it will be of order the number of $e$-foldings) but the combination $\epsilon \times \log$ is still small.  In this approximation, the relevant parts of the cubic solid Lagrangian for calculating the leading order contribution to the squeezed three-point function are
\begin{align}
\mathcal{L}_{\gamma \gamma \gamma} = &  \, \mpl^2\, a^3   H^2\frac{F_Y}{F}  \Big\{-\sfrac{1}{9}\gamma_{ij}\gamma_{jk}\gamma_{ki}\Big\}  \, , 
\label{cubicgamma}
\end{align}
\begin{align}
\mathcal{L}_{\pi \gamma \gamma} = &  \, \mpl^2\, a^3   H^2\frac{F_Y}{F}  \Big\{-\sfrac{2}{9}(\di \cdot \pi)\gamma_{ij} \gamma_{ji} +\sfrac{2}{3}\gamma_{ij} \gamma_{jk} \di^k\pi^i \Big\} \,. \label{cubicggp}
\end{align}
\noindent which yield the soft limits
\begin{align}
\langle \zeta_{\vec q \rightarrow 0} \gamma^{s}_{\vec k} \gamma^{s'}_{-\vec k} \rangle'=  & \, \frac{16}{9}\, \frac{F_Y}{F}\,P_{\zeta}(q) P_\gamma (k)\,\log\left(\frac{k}{a H}\right)\,  \nonumber\\ 
&\times \big(\epsilon^s_{ij}\epsilon^{s'}_{ji} - 3 \hat q ^i \epsilon^s_{ij}\epsilon^{s'}_{jk}\hat q ^k \big) \; ,
\end{align}
\begin{align}
\langle \pi_{\lambda, \vec q \rightarrow 0} \gamma^s_{\vec k} \gamma^{s'}_{-\vec k} \rangle' = &-\frac{8}{9} \,iq\,\frac{F_Y}{F}\,P_{T}(q) P_\gamma (k)\,\log\left(\frac{k}{a H}\right) \nonumber \\
& \times \varepsilon^i_\lambda \{ \epsilon^s_{ij} \epsilon^{s'}_{jk} + \epsilon^{s'}_{ij} \epsilon^{s}_{jk} \} \hat q ^k \; ,
\end{align}
\begin{align}
\langle \gamma^s_{\vec q \rightarrow 0} \gamma^{s'}_{\vec k} \gamma^{s''}_{-\vec k} \rangle' = & \, \frac{8}{9} \,\frac{F_Y}{F}\,P_{\gamma}(q) P_\gamma (k)\,\log\left(\frac{k}{a H}\right)  \nonumber\\ & \times \epsilon^s_{ij} \epsilon^{s'}_{jk} \epsilon^{s''}_{ki}\, .
\end{align}

\noindent where the first two of these relations can be checked to satisfy the obvious generalization of the sum rule \eqref{sum rule} when the appropriate trace over tensor polarizations is taken.

Note that the extra powers in \eqref{correctedpower} can in principle give an $O(1)$ correction to the power spectrum.  In this limit, the combination $\epsilon \times \log$ is of order one and formally perturbation theory breaks down, unless we are able to resum all the terms with arbitrary powers of $\epsilon \times \log$.  It would be interesting to understand and to formalize this along the lines of standard renormalization group techniques.


\end{document}